# *p*-type $Bi_2Se_3$ for topological insulator and low temperature thermoelectric applications

Y.S. Hor[1], A. Richardella[2], P. Roushan[2], Y. Xia[2], J.G. Checkelsky[2],

A. Yazdani[2], M.Z. Hasan[2], N.P. Ong[2], and R.J. Cava[1]

[1]Department of Chemistry, Princeton University

[2]Department of Physics, Princeton University

**Abstract**

The growth and elementary properties of *p*-type $Bi_2Se_3$ single crystals are reported. Based on a hypothesis about the defect chemistry of $Bi_2Se_3$, the *p*-type behavior has been induced through low level substitutions (1% or less) of Ca for Bi. Scanning tunneling microscopy is employed to image the defects and establish their charge. Tunneling and angle resolved photoemission spectra show that the Fermi level has been lowered into the valence band by about 400 meV in $Bi_{1.98}Ca_{0.02}Se_3$ relative to the *n*-type material. *p*-type single crystals with *ab* plane Seebeck coefficients of +180 µV/K at room temperature are reported. These crystals show a giant anomalous peak in the Seebeck coefficient at low temperatures, reaching +120 µVK$^{-1}$ at 7 K, giving them a high thermoelectric power factor at low temperatures. In addition to its interesting thermoelectric properties, *p*-type $Bi_2Se_3$ is of substantial interest for studies of technologies and phenomena proposed for topological insulators.

**Introduction**

$Bi_2Se_3$ is one of the binary end-members of the $(Bi,Sb)_2(Te,Se)_3$ family of thermoelectric materials. Decades of work in the chemistry, physics, and processing of these materials has led to complex formulations of compounds and microstructures optimized for use as thermoelectrics under various conditions (see e.g. refs. 1 and 2). Due to the superiority of $Bi_2Te_3$-based materials for these applications, $Bi_2Se_3$, while well studied, has not been subject to the same degree of intensive research as has its heavier mass analog. One of the major issues for $Bi_2Se_3$-based thermoelectrics has been the difficulty in making the material *p*-type. Unlike $Bi_2Te_3$, which can simply be made *n*- or *p*-type through variation of the Bi:Te ratio, the defect chemistry in $Bi_2Se_3$ is dominated by charged selenium vacancies, which act as electron donors [3], resulting in *n*-type behavior for virtually all of the reported transport studies (see e.g. refs. 3-11). *p*-type behavior for pure $Bi_2Se_3$ was reported in an early study [9], but never since. Modern studies have shown that *p*-type behavior is possible when beginning with an *n*-type host material in the $(Bi_{2-x}Sb_x)Se_3$ solid solution with x = 0.4, and then doping that composition with small amounts of Pb to create a quaternary *p*-type material [10].

The present study is motivated by the desire to find a chemically less complex *p*-type $Bi_2Se_3$-based material to address the recent emergence of $Bi_2Se_3$ as one of the prime candidates for the study of topological surface states (see e.g. refs. 12-16), and for thermoelectric applications. The character and stability of the surface states in $Bi_2Se_3$ at room temperature has motivated the suggestion that they may be useful for quantum computing applications if the Fermi level can be lowered to the Dirac point through *p*-type doping of the normally *n*-type compound [15,17,18]. Therefore high quality *p*-type crystals are important for both fundamental and applied research on $Bi_2Se_3$.

Unlike the case for $Bi_2Te_3$, where the chemical similarity of Bi and Te leads to antisite defects as the primary source of carrier doping in binary compounds, there is little tendency for Bi/Se mixing in $Bi_2Se_3$, and the primary structural defect giving rise to electron doping is doubly charged selenium vacancies $V_{Se}^{\bullet\bullet}$ [3]:

$$Se_{Se} \rightarrow V_{Se}^{\bullet\bullet} + Se(g) + 2e'$$

Ordinarily, one compensates for the presence of donors in this chemical family through doping with Pb on the Bi site, as Pb has one fewer electron than Bi. This substitution does not, however, lead to the formation of *p*-type material for $Bi_2Se_3$. We hypothesize that this is due to the fact that Pb in $Bi_2Se_3$ is ambipolar, much as is the case for Cu [11]. This indicates that a more ionic substitution on the Bi site may be required for hole-doping of $Bi_2Se_3$, suggesting the use of Ca substitution, with the defect reaction:

$$2Ca \xrightarrow{Bi_2Se_3} 2Ca_{Bi}' + 2h^{\bullet}$$

with Ca substitution for Bi creating a negatively charged defect ($Ca_{Bi}'$) that in turn generates holes ($h^{\bullet}$) to compensate the electrons created by the Se vacancies. Our results, described below, support this scenario as a good representation of the defect chemistry in this compound.

**Experimental**

The single crystals of $Bi_{2-x}Ca_xSe_3$ were grown via a process of two-step melting, starting with mixtures of high purity elements (Bi, 99.999 %, Se, 99.999 %, Ca, 99.8 %). First, stoichiometric mixtures of Bi and Se were melted in evacuated quartz ampoules at 800 ºC for 16 hours. The melts were then stirred before being allowed to solidify by air-quenching to room temperature. Second, the stoichiometric amount of Ca was added in the form of pieces, using care to avoid direct contact of the added Ca with the quartz. The materials were then heated in evacuated quartz ampoules at 400 ºC for 16 hours followed by 800 ºC for a day. The crystal growth process involved cooling from 800 to 550 ºC over a period of 24 hours and then annealing at 550 ºC for 3 days. The crystals were then furnace cooled to room temperature. They were cleaved very easily along the basal plane, and were cut into approximately 1.0 × 1.0 × 6.0 mm$^3$ rectangular bar samples for the thermal and electronic transport measurements. Resistivity, Seebeck coefficient, and thermal conductivity measurements were performed in a Quantum Design PPMS, using the standard four-probe technique, with silver paste cured at room temperature used for the contacts. Hall Effect measurements to determine carrier concentrations were performed at 1.5 K in a home-built apparatus. In all cases, the electric- and thermal-currents were applied in the basal plane (*ab* plane in the hexagonal setting of the rhombohedral cell) of the crystals. In order to probe the electronic states of native and Ca related defects, Bi$_2$Se$_3$ and Bi$_{1.98}$Ca$_{0.02}$Se$_3$ samples were studied in a cryogenic scanning tunneling microscope (STM) at 4.2 K. The samples were cleaved *in situ* in ultrahigh vacuum to expose a pristine surface. The surfaces of the *p*-type crystals are fully stable in the STM experiments. High-resolution angle resolved photoemission spectroscopy (ARPES) measurements were performed using 22-40 eV photons at Beamline-12 of the Advanced Light Source (Berkeley Laboratory) and on beamline 5 at the Stanford Synchrotron Radiation Laboratory. The energy and momentum resolutions were 15 meV and 2 % of the surface Brillouin Zone, respectively, obtained using a Scienta analyzer. The samples were cleaved at 10 K under pressures of less than 5 × 10$^{-11}$ torr, resulting in shiny flat surfaces. The spectra employed are those closest to the time of cleavage (about 10 minutes), which are the most representative of the bulk.

**Results**

The rhombohedral crystal structure of Bi$_2$Se$_3$ consists of hexagonal planes of Bi and Se stacked on top of each other along the [001] crystallographic direction (hexagonal setting), with the atomic order: Se(1)-Bi-Se(2)-Bi-Se(1), where (1) and (2) are refer to different lattice positions [19]. The unit cell consists of three of these units stacked on top of each other with weak van-der-Waals bonds between Se(1)-Se(1) layers, making the (001) plane the natural cleavage plane. Scanning tunneling spectroscopy (dI/dV) was performed to measure the density of states. The spectra of $Bi_2Se_3$ were $n$-type, as expected, due to the presence of Se vacancies with the Fermi energy near the conduction band, consistent with previous STM results [20]. In contrast, a clear shift in the position of the Fermi level towards the valance band occurs in the $Bi_{1.98}Ca_{0.02}Se_3$ sample, as shown in Fig. 1(a). The finite density of states inside the gap is due to surface states, for which novel topological properties have been predicted [12-18].

We were able to identify various defects and the sign of their charge state from the STM topographies of the filled states and unoccupied states. The STM topographies of the native $Bi_2Se_3$ (001) surface are dominated by one type of defect, which appears as a bright triangular protrusion in the topographies of the unoccupied states (Fig. 1(b)). On average, these defects are about 40 Å across, but vary in size between defects indicating that they are located in various layers beneath the surface. Given that no other defects were observed, we attribute these triangular defects to Se vacancies. Figs. 1(c) and (d) show the topography of the unoccupied and occupied states of the Ca-doped $Bi_2Se_3$, respectively. Comparison between 1(b) and (c) makes it clear that the density of triangular defects has been reduced significantly in the Ca-doped samples. In addition, the STM topography of the $Bi_{1.98}Ca_{0.02}Se_3$ (001) surface shows distinct defects that were not present in $Bi_2Se_3$ samples. In topographic images of the empty states, the shape of these three-fold symmetric defects resembles a cloverleaf (Fig. 1(c)). Based on their two distinct spatial extents we conclude that they are located at two different crystallographic positions; most likely the smaller one is located in a layer nearer the surface and other one in a layer deeper beneath the surface. Because these defects occur only in Ca-doped samples, we identify them as Ca-related defects.

The charge state of the defects can be inferred by observing the bending of the host bands caused by the coulomb field surrounding a charged defect. A positively charged defect lowers the electronic energy level in its neighboring region, leading to a depression area in the STM topography of the filled states, and an enhancement in the topography of the unoccupied states. We expect this to be observed in imaging the ionized donors, and the opposite effect for imaging negatively charged defects, such as acceptors. The cloverleaf shape defects are surrounded by a region of enhancement in the topography of the filled states (Fig. 1(d)), implying that they are negatively charged, consistent with a Ca acceptor. A comparison between the triangular defect close to the center of the image in Fig. 1(c) and (d) shows a region of depression in the topography of the filled states (c), and hence implies the presence of a positively charged defect. This observation is expected for Se vacancies, which are known to be electron donors and in their ionized state be positively charged. Thus the STM data support the defect chemistry model described above for Ca-doped $Bi_2Se_3$, with the added observation that the concentration of defects in general is lower in the Ca-doped samples than for native $Bi_2Se_3$.

The ARPES data showing valence band energy dispersion curves in the vicinity of the $\Gamma$ point of the Brillouin Zone for *n*- and *p*- type $Bi_2Se_3$ are presented in Figs. 2a and 2b. For the *n*-type crystal, the valence band is clearly observed below $E_F$, as is a small pocket of the conduction band near the $\Gamma$ point. For the *p*-type crystal, only the valence band is observed. This chemical potential shift indicates that the Ca doped crystal is hole doped relative to the *n*-type $Bi_2Se_3$ crystal. In the *p*-type doped samples, the surface state bands are not observed, indicating that the Fermi level is below the energy of the Dirac point. Judging from the position of the strongest valence bands, the energy dispersion curves through the Γ point show that the chemical potential in *p*-type $Bi_{1.98}Ca_{0.02}Se_3$ is shifted by approximately 400 meV relative to the *n*-type crystal (Fig. 2c). Lower hole doping levels, which would be obtained by tuning the Ca concentration, are expected to lead to materials whose chemical potential is very close to the Dirac point, and therefore of interest in fundamental and applied studies of the topological surface states [15].

The 2–300 K resistivities in the *ab* plane for undoped $Bi_2Se_3$ and $Bi_{2-x}Ca_xSe_3$ crystals for x = 0.005, 0.02 and 0.05 are shown in Fig. 3. All show the weakly metallic resistivities commonly seen in high carrier concentration small band gap semiconductors, with resistivities in the 0.3 to 1.5 mΩ-cm range at temperatures near 10 K. The lightly doped x = 0.005 material is *p*-type with a carrier concentration at room temperature determined by Hall effect measurements (lower inset, Fig. 3) to be approximately $1 \times 10^{19}$ $cm^{-3}$. This crystal has a resistivity ratio, $\rho_{300}/\rho_{4.2}$ of about 3, as does the undoped *n*-type crystal. The upper inset shows the low temperature resistivity region in more detail for *n*- and *p*-type crystals. The decreasing resistivity seen on cooling is arrested at these low temperatures and rises slightly below 20-30 K, suggesting that carriers have been frozen out in this temperature regime.

The corresponding *ab* plane Seebeck coefficients are shown in Fig. 4. The room temperature Seebeck coefficient for the *n*-type undoped $Bi_2Se_3$ crystal, -190 $\mu VK^{-1}$, is large compared to those usually reported for this material, which are typically in the -50 to -100 $\mu VK^{-1}$ range (3-11). The carrier concentration for this *n*-type $Bi_2Se_3$ is $8 \times 10^{17}$ $cm^{-3}$ (lower inset, Fig. 2). The magnitude of the Seebeck coefficient decreases smoothly with temperature until around 15 K (inset Fig. 4), where it becomes distinctly more negative in a small peak with an onset temperature roughly corresponding to that of the upturn in the resistivity. A similar peak in Seebeck coefficient has been observed previously in this temperature range in *n*-type $Bi_2Se_3$ [3]. The *p*-type materials show similar but more unusual behavior. The room temperature Seebeck coefficient for the lightly doped *p*-type material is again very high, reaching approximately +180 $\mu VK^{-1}$. This value is larger than what is observed in the quaternary *p*-type $(Bi,Sb,Pb)_2Se_3$ materials [10] and remains large and *p*-type over the full temperature range of measurement. Crystals with Ca-doping levels as low as $Bi_{1.9975}Ca_{0.0025}Se_3$ are *p*-type. A dramatic low temperature peak in the Seebeck coefficient (inset Fig. 3) is seen for all *p*-type compositions studied. The low temperature Seebeck coefficient peaks of -60 $\mu VK^{-1}$ and +120 $\mu VK^{-1}$ for pure and doped $Bi_2Se_3$ at 7 K represent very high values when compared to other small band gap semiconductors at this low temperature. Interestingly, similarly anomalous low temperature peaks have been observed in $Bi_{1-x}Sb_x$ alloys, another material in the vicinity of a Dirac point [12-14], in the critical "zero gap" composition region near x = 0.07 [21].

The total thermal conductivities ($\kappa$) in the *ab* plane are shown in Fig. 5. They display the behavior typical of high quality crystals, increasing with decreasing temperature as the phonon contribution to $\kappa$ grows, until low temperatures, where the phonon mean free path becomes comparable to intra-defect distances. The low temperature regime for *n*- and *p*-type crystals is shown in the inset. The thermal conductivity for the *n*-type crystal reaches the relatively high value of about 100 $WK^{-1}m^{-1}$ at 10 K, and all the Ca-doped materials show a somewhat lower maximum thermal conductivity, in the 40-50 $WK^{-1}m^{-1}$ range.

The efficiency of a material for thermoelectric cooling or power generation at temperature T is described by the thermoelectric figure of merit ZT [22]. The materials parameter is $Z = (S^2/\rho)(1/\kappa)$, where the power factor, defined as the square of the Seebeck coefficient S divided by the electrical resistivity, has been separated from the total thermal conductivity κ. Due to the very high thermal conductivities of the single crystals, the $Bi_2Se_3$-based *n*- and *p*-type materials are of limited use as practical thermoelectric materials in single crystal form, especially at low temperatures. The resulting low values of the thermoelectric figure of merit are shown in the inset to Fig. 6. The Seebeck coefficients and electrical resistivities do, however, suggest that these materials may have potential for use as thermoelectrics in the very low temperature regime. To illustrate this fact we have plotted the temperature dependences of the power factors in the main panel of Fig. 6. The anomalous low temperature peaks in all the materials, especially for the *p*-type material, where the power factor at 7 K is higher than it is at 300 K, and comparable to that of the *n*-type material at 300 K, are particularly intriguing for possible low temperature thermoelectric applications. The underlying origin of these peaks, and also the more subtle anomalies seen in the transport properties in the 30-40 K range in all materials, is not presently known.

**Conclusions**

Through consideration of the defect chemistry of $Bi_2Se_3$, we have identified Ca as a dopant that when present in sub-percent quantities results in the formation of a *p*-type material. The use of *p*-type $Bi_2Se_3$ for the fundamental study of topological surface states and for devices based on these states for quantum computing will be of considerable future interest. The dramatic low temperature peaks found in the Seebeck coefficients in $Bi_{1-x}Sb_x$ and $Bi_2Se_3$, materials that have strong spin orbit coupling and nearby Dirac points [12-16,19], suggests that these characteristics may be related. In addition, processing of the microstructure of *n*- and *p*-type $Bi_2Se_3$ to reduce the thermal conductivity to the < 1 $WK^{-1}m^{-1}$ range typically observed for polycrystalline materials in this family, if it can be performed while maintaining the power factors observed in the current work, promises the possibility of low temperature thermoelectric applications. Finally, the development of more sophisticated crystal growth methods for lightly-doped *p*-type $Bi_{2-x}Ca_xSe_3$ crystals will be of interest for the study of topological surface states.

**Acknowledgements**

This work was supported by the AFOSR, grant FAA9550-06-1-0530, and also by the NSF MRSEC program, grant DMR-0819860.

**Figures**

**Fig. 1.** (Color on line) (a) Spatially averaged density of states (dI/dV) measurements showing the shift of $E_F$ between *n*-type $Bi_2Se_3$ and *p*-type $Bi_2Ca_{0.02}Se_3$. (b) STM topography of the empty states of $Bi_2Se_3$ ($V_B$ = +1 V, and I = 10 pA) showing triangular shaped defects observed at various intensities, indicating they are located in different layers beneath the surface. (c) Topography of the empty states of $Bi_{1.98}Ca_{0.02}Se_3$ ($V_B$ = +2.0 V, and I = 10 pA) showing clover leaf looking defects and a substantial reduction in the density of the triangular defects, which dominated the undoped samples. (d) Topography of the filled states over the same area as in (c) ($V_B$ = -1.0 V, and I = 10 pA). The area around the triangular defect near center shows a depression around it in the filled states, demonstrating it is positively charged. In contrast, the Ca-related defects exhibit an area of enhancement around them, indicating they are negatively charged. All STM topographies are 500 Å by 500 Å.

**Fig. 2.** (color on line) (a) The valence band structure of *n*-type $Bi_2Se_3$ measured by ARPES near the Fermi energy and the Γ point of the Brillouin Zone. The chemical potential (the Fermi level) is about 0.4 eV above the top of the valence band, about 0.3 eV above the surface Dirac point (15). (b) The band structure of a Ca-doped $Bi_2Se_3$ crystal ($Bi_{1.98}Ca_{0.02}Se_3$) measured under similar conditions. The chemical potential is within 30 meV of the valence band. For (a) and (b) the Γ point is 0.0 on the horizontal axes and the band dispersions are shown in the ±M directions. (c) The momentum-selective density of states (DOS) near the Γ point, compared for the *n*- and *p*-type crystals, covering a large binding energy range. A large shift of the chemical potential with Ca doping can be traced by considering the shift of DOS peaks in the valence band. The drop of the chemical potential by about 380 meV confirms the hole doped nature of the Ca doped samples.

**Fig. 3.** (color on line) Temperature dependent resistivity between 2 and 300 K in the *ab* plane of single crystals of $Bi_2Se_3$ and Ca-doped variants. The stoichiometric material is *n*-type and the Ca-doped materials are all *p*-type. The upper inset shows the resistivities in the low temperature regime. The lower inset shows the Hall effect data employed to determine the carrier concentrations for undoped $Bi_2Se_3$ and one Ca-doped crystal.

**Fig. 4.** (color on line) Temperature dependent Seebeck coefficients between 2 and 300 K in the *ab* plane of single crystals of $Bi_2Se_3$ and Ca-doped variants. The inset shows the low temperature region for both pure and lightly Ca-doped $Bi_2Se_3$. The undoped $Bi_2Se_3$ is *n*-type, with a carrier concentration of $8 \times 10^{17}$ cm$^{-3}$, and the Ca-doped material with x = 0.005 is *p*-type with a carrier concentration of $1 \times 10^{19}$ cm$^{-3}$.

**Fig. 5.** (color on line) Temperature dependent total thermal conductivities between 2 and 300 K in the *ab* plane of single crystals of $Bi_2Se_3$ and Ca-doped variants. The inset shows the low temperature region for both pure and lightly Ca-doped $Bi_2Se_3$.

**Fig. 6.** (color on line) Temperature dependent thermoelectric power factor between 2 and 300 K in the *ab* plane of single crystals of $Bi_2Se_3$ and Ca-doped variants. All materials show a peak at low temperatures but the peak in the *p*-type variant with 1 % Ca doping ($Bi_{2-x}Ca_xSe_3$ with x = 0.02) is particularly dramatic. The inset shows the thermoelectric figure of merit in the same temperature range.

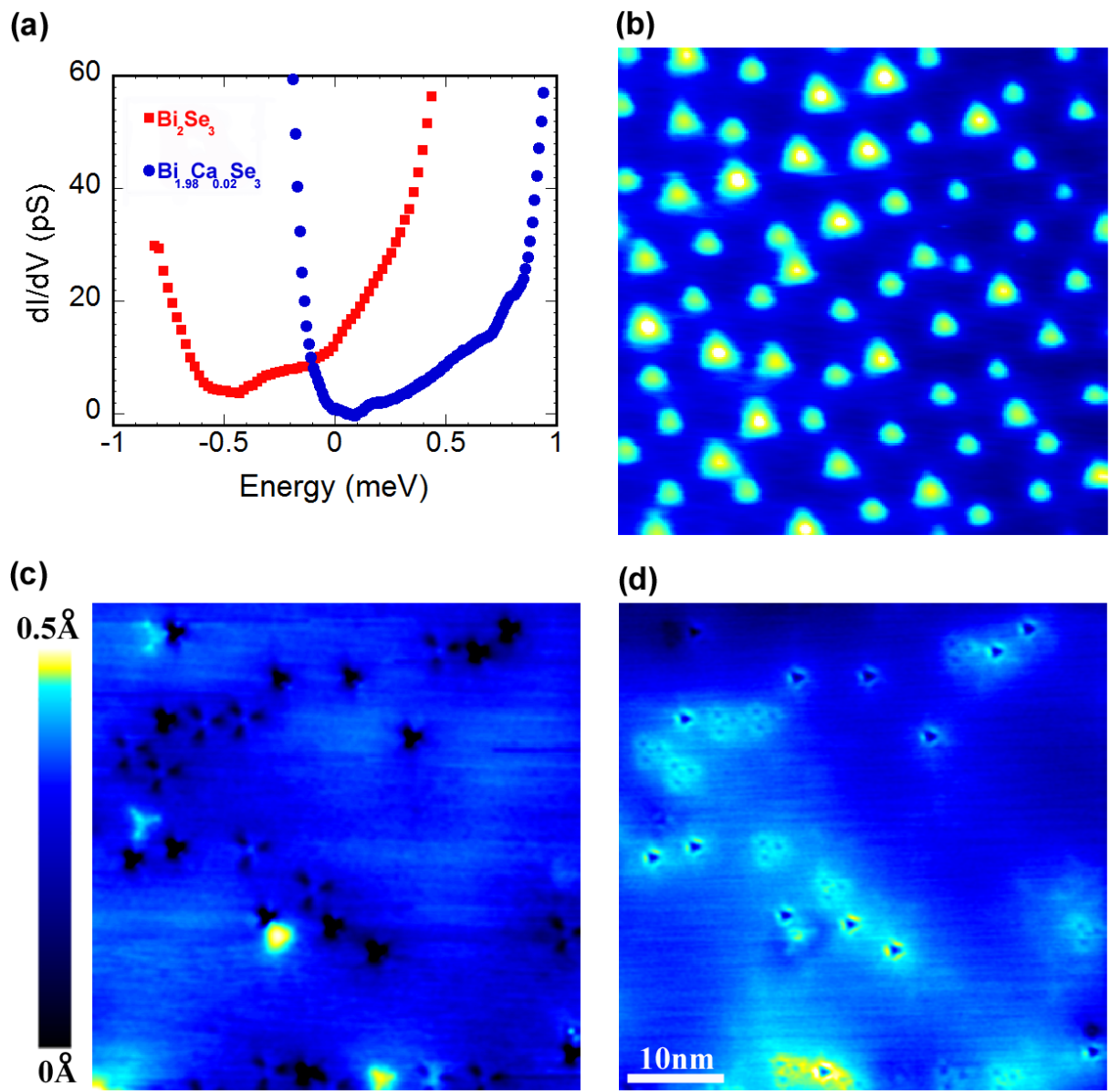

**Figure 1**

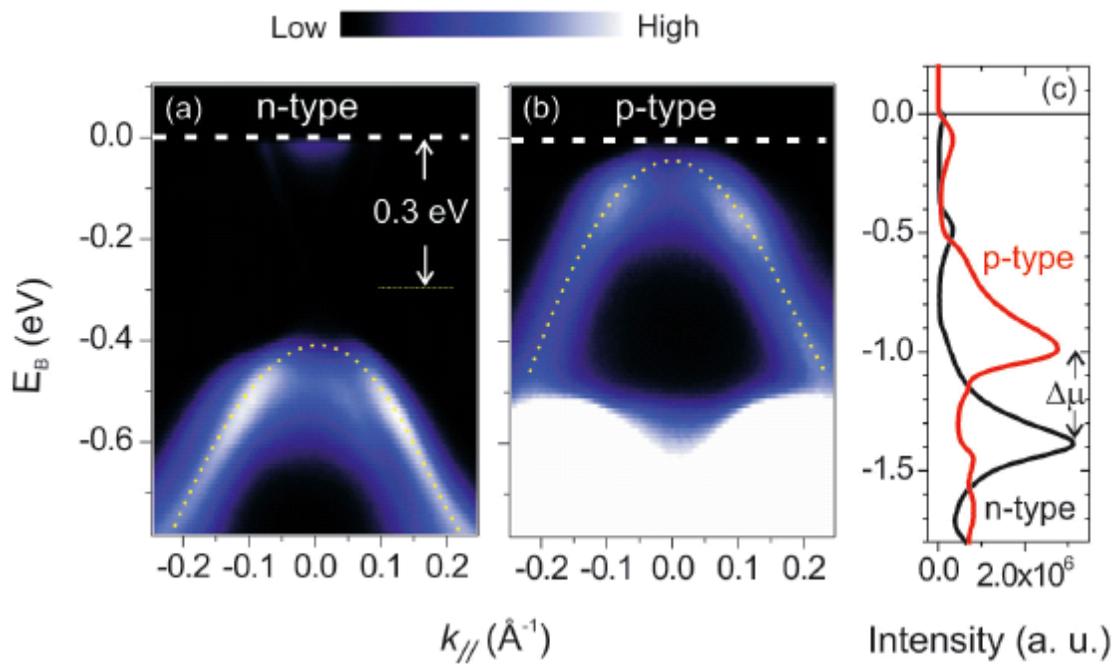

**Figure 2**

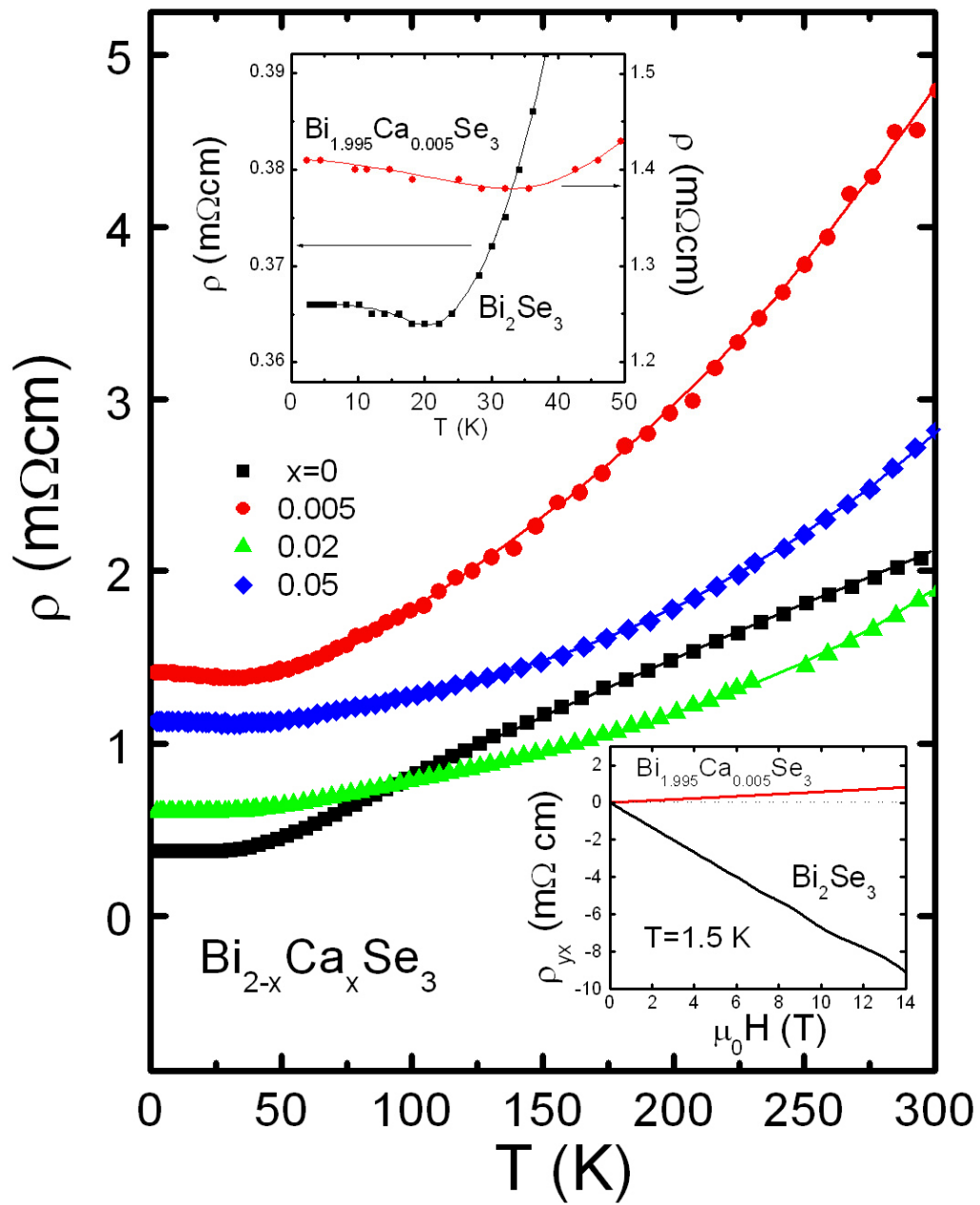

**Figure 3**

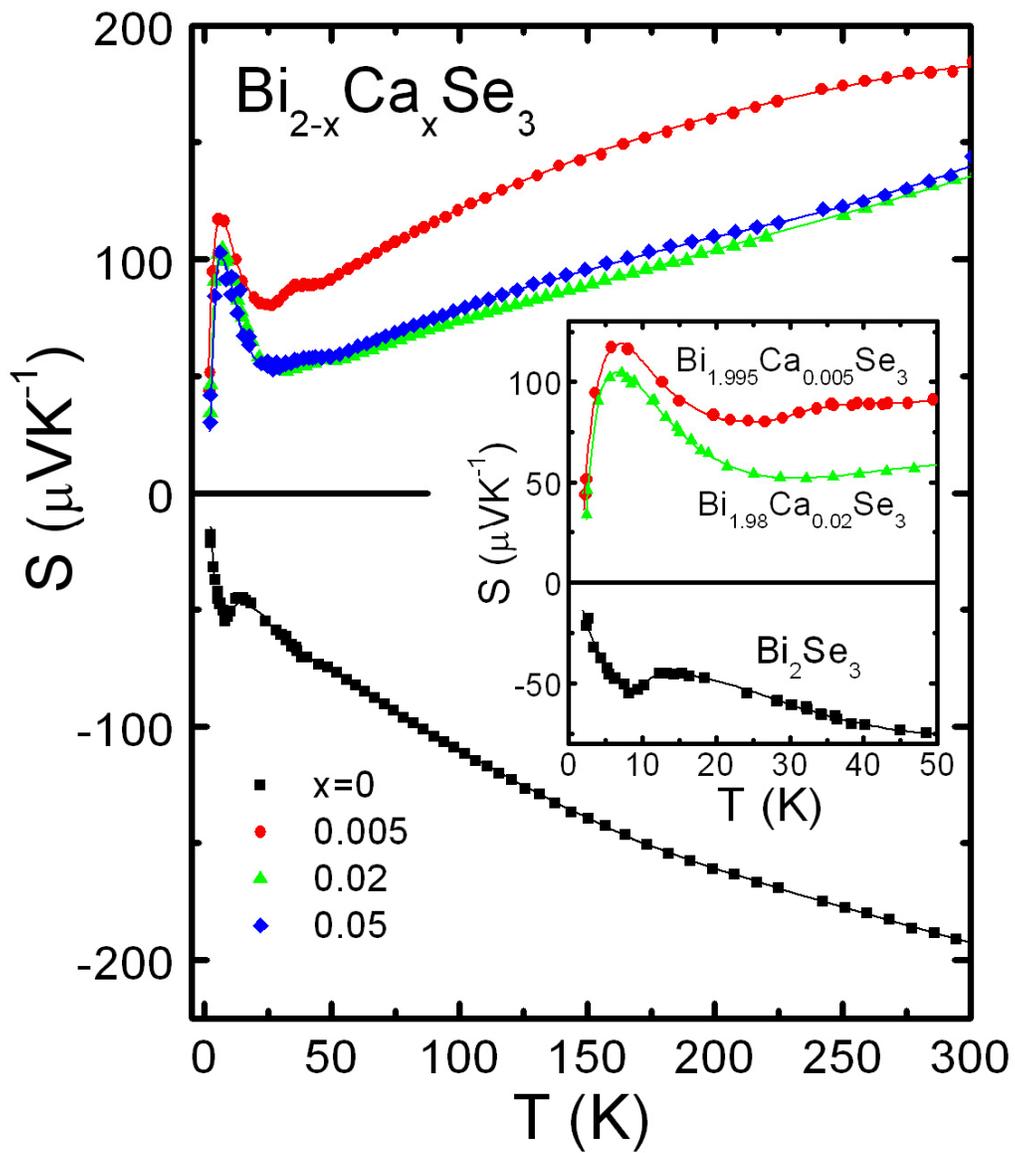

**Figure 4**

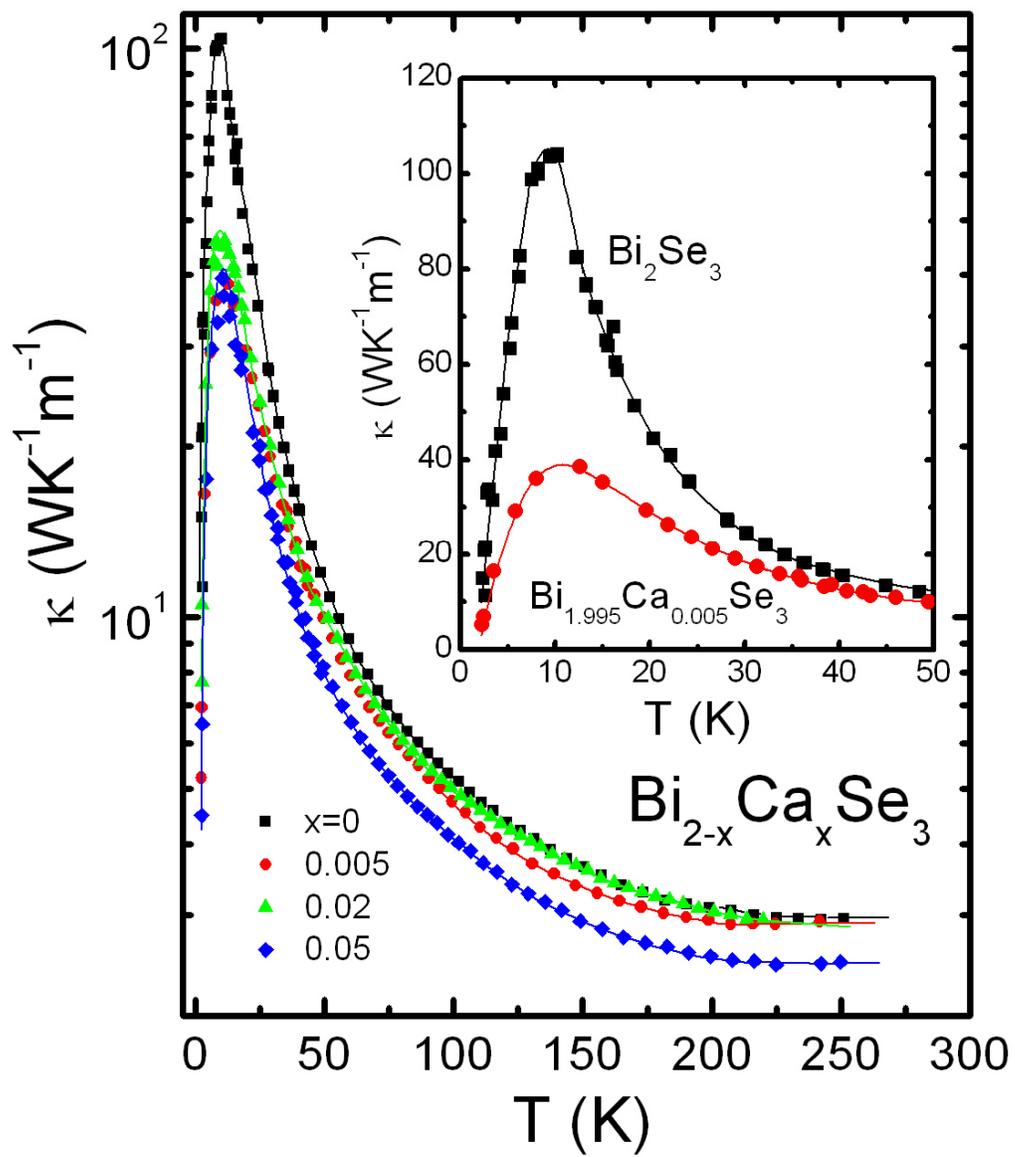

**Figure 5**

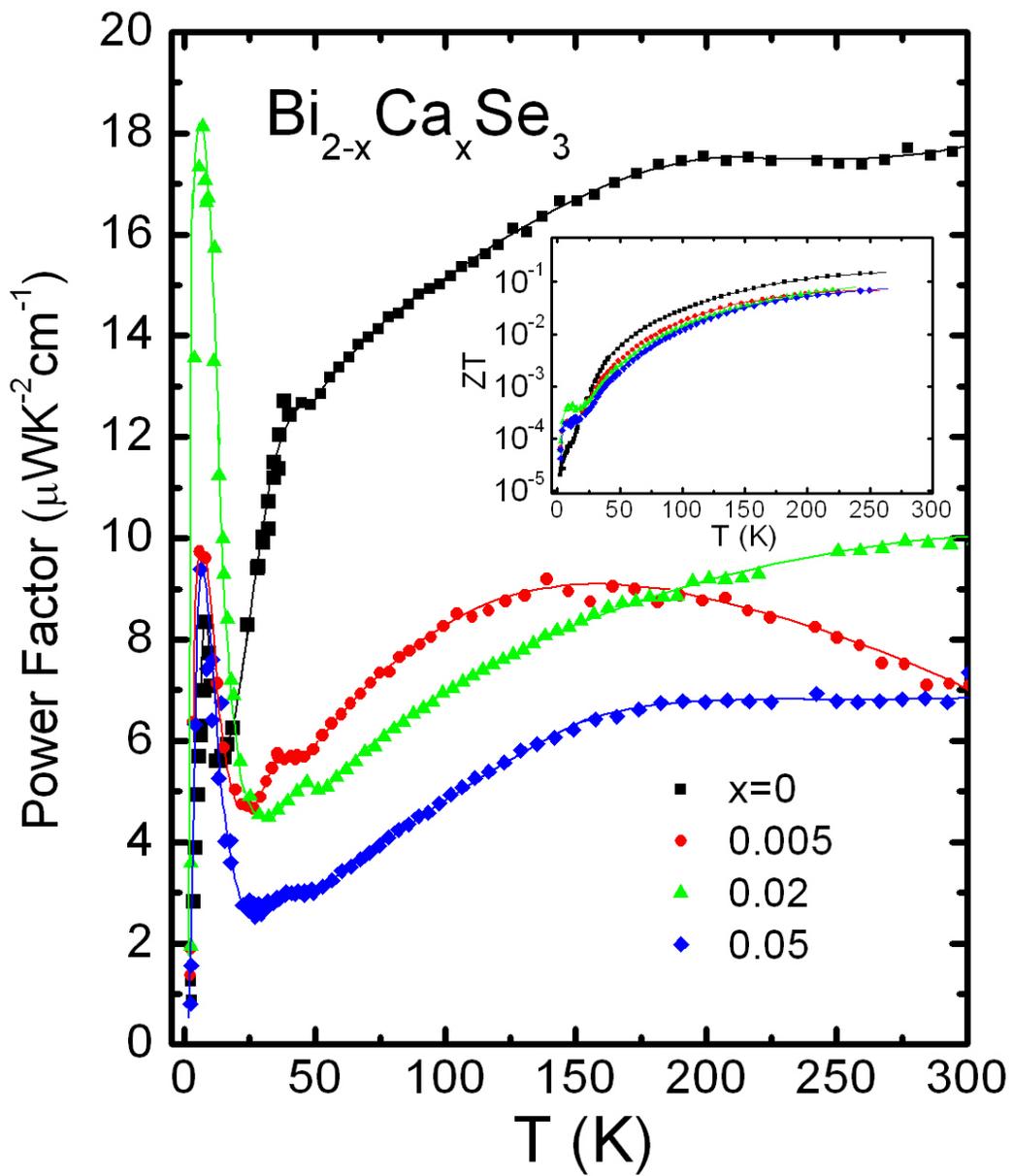

**Figure 6**